\documentstyle[prl,aps,multicol,epsfig]{revtex}
\renewcommand{\narrowtext}{\begin{multicols}{2}
\global\columnwidth20.5pc} 
\renewcommand{\widetext}{\end{multicols}
\global\columnwidth42.5pc} \multicolsep = 8pt plus 4pt minus 3pt

\begin{document}
\draft
\title{$3e$ tunneling processes in a superconducting single-electron 
tunneling transistor}
\author{P. Hadley, E. Delvigne, E. H. Visscher, S. L\"ahteenm\"aki, 
and J. E. Mooij}
\address{ Applied Physics and DIMES, Delft University of Technology\\
Lorentzweg 1, 2628 CJ Delft, The Netherlands}
\date{\today}
\maketitle
\begin{abstract}
A current due to a tunneling event that involves three times the charge 
of an electron was observed in the current - voltage characteristics of 
a superconducting single-electron tunneling transistor. In this tunnel 
event, a Cooper pair tunnels through one tunnel barrier simultaneously 
with a quasiparticle that tunnels through a second tunnel barrier which 
is about 0.5 $\mu $m distant from the first tunnel barrier. This current 
was observed in a bias regime where current flow due to sequential 
quasiparticle tunneling is forbidden due to the Coulomb blockade.
\end{abstract}

\narrowtext

A superconducting single-electron tunneling (SET) transistor 
consists of a small superconducting island that is coupled to 
three leads, a gate and two output leads.\cite{likharev87} The 
two output leads 
are connected to the island by tunnel junctions and the gate is 
capacitively coupled to the island. The quantum nature of this 
device is manifested in the periodic modulation of the current 
that flows through the output leads as the charge on the gate is 
varied. One modulation period corresponds to adding one electron 
charge $e$ to the island. By monitoring the current, one can make 
very sensitive measurements of the charge at the gate. The charge 
sensitivity of a SET transistor in the superconducting state is 
better than the charge sensitivity of a SET transistor in the 
normal state which makes the superconducting SET transistor the 
most sensitive device now available for measuring charge. 
\cite{korotkov96,schoelkopf98}

The characteristics of a superconducting SET transistor 
depend on the relative magnitudes of three energies: 
the charging energy $E_{C}$, the Josephson energy $E_{J}$, 
and the superconducting gap $\Delta $. \cite{tinkham96} 
The charging energy is the energy associated with charging 
the island with a single electron charge, 
$E_C = e^2/\left( 2C_{\Sigma}\right) $. Here 
$C_{\Sigma}$ is the total capacitance of the island. 
The Josephson energy is related to the junction critical 
current $I_c$, $E_J = \frac{\hbar I_c}{2e}$, and the 
superconducting gap can be seen as the addition energy that is 
required for a superconducting island to have an odd number of 
electrons rather than an even number of electrons.\cite{averin92} 
In devices with large junctions, the Josephson energy is much 
larger than the charging energy, $E_{J} >> E_{C}$, and a 
supercurrent is observed. As the junctions are made smaller, 
$E_{J}$ decreases while $E_{C}$ increases. When $E_{J}\approx 
E_{C}$, the supercurrent can be modulated by applying a voltage 
to the gate, while for $E_{J}<<E_{C}$ the supercurrent is 
suppressed. If $\Delta >E_{C}>E_{J}$ parity effects are 
observed.\cite{touminen92} It is then possible to determine 
if the number of electrons on the island is an odd or even 
number. In the present experiment $\Delta \approx E_{C}>>E_{J}$ 
and no supercurrent was observed.

The SET transistor studied consisted of two Al/AlO$_{x}$/Al 
tunnel junctions that were fabricated by shadow evaporation. 
The two junction capacitances were $C_{1}=1.78 \times 10^{-16}$ F 
and $C_{2} = 2.10 \times 10^{-16}$ F, the gate capacitance 
was $C_{g}=1.07 \times 10^{-18}$ F, the total resistance 
of the device was $R_{1}+R_{2}=1.7\times 10^{6}$ $\Omega$,  
the superconducting gap was $\Delta = 203$ $\mu$eV, and the 
charging energy was $E_{C}=206$ $\mu$eV. 
Under normal operating conditions, the current that flows 
through a superconducting SET transistor is primarily due 
to the sequential tunneling of normal quasiparticles. 
However, at low bias voltages, the tunneling of individual 
quasiparticles is suppressed by a combination of the Coulomb 
blockade and the absence of states in the superconducting gap. 
At these low bias voltages, other transport mechanisms can be 
observed such as cotunneling,\cite{averin91} the Josephson - 
quasiparticle cycle,\cite{fulton89,manninen97} Andreev reflection,
\cite{fitzgerald98} the resonant tunneling of Cooper pairs,
\cite{haviland94} and singularity matching. 
\cite{nakamura98} Here we report the experimental observation 
of a current that flows due to the simultaneous tunneling of a 
Cooper pair and a quasiparticle. The Cooper pair and the 
quasiparticle simultaneously tunnel through two different 
tunnel barriers that are spaced about 0.5 $\mu$m from each 
other.

\begin{figure} [htb]
\epsfig{figure=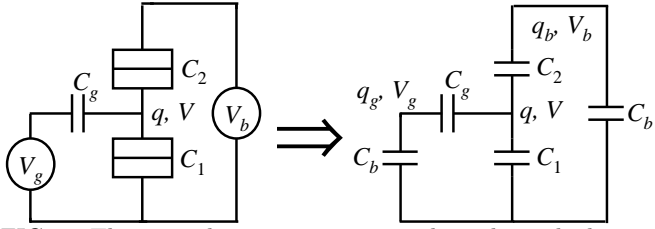, width=20.5pc, clip=true}
\caption{The equivalent capacitor network used to calculate 
the electrostatic energy of a SET transistor.}
\label{circuit}
\end{figure}

The thresholds for the various tunnel events that occur in a 
SET transistor can be determined by examining the electrostatic 
energy of the circuit. To calculate the change in electrostatic 
energy when an electron tunnels, one can treat the circuit as a 
network of capacitors. It is convenient to also treat the voltage 
sources as capacitors with very large capacitances.\cite{devoret97} 
At the end of 
the calculation the limit of very large capacitance for the voltage 
sources is taken. Figure\ \ref{circuit} shows the equivalent 
capacitor network for an asymmetrically biased SET transistor. 
The electrostatic energy of this network of capacitors is the 
sum of the electrostatic energies of the capacitors, 

\begin{eqnarray}
E&=&\frac{1}{2}C_1V^2+\frac{1}{2}C_2\left( V-V_b\right) ^2+
\frac{1}{2}C_g\left( V-V_{g}\right) ^2 \nonumber \\
& &+\frac{1}{2}C_bV_{b}^2+
\frac{1}{2}C_bV_{g}^2.
\label{eqn1}
\end{eqnarray}

\begin{figure} [htb]
\epsfig{figure=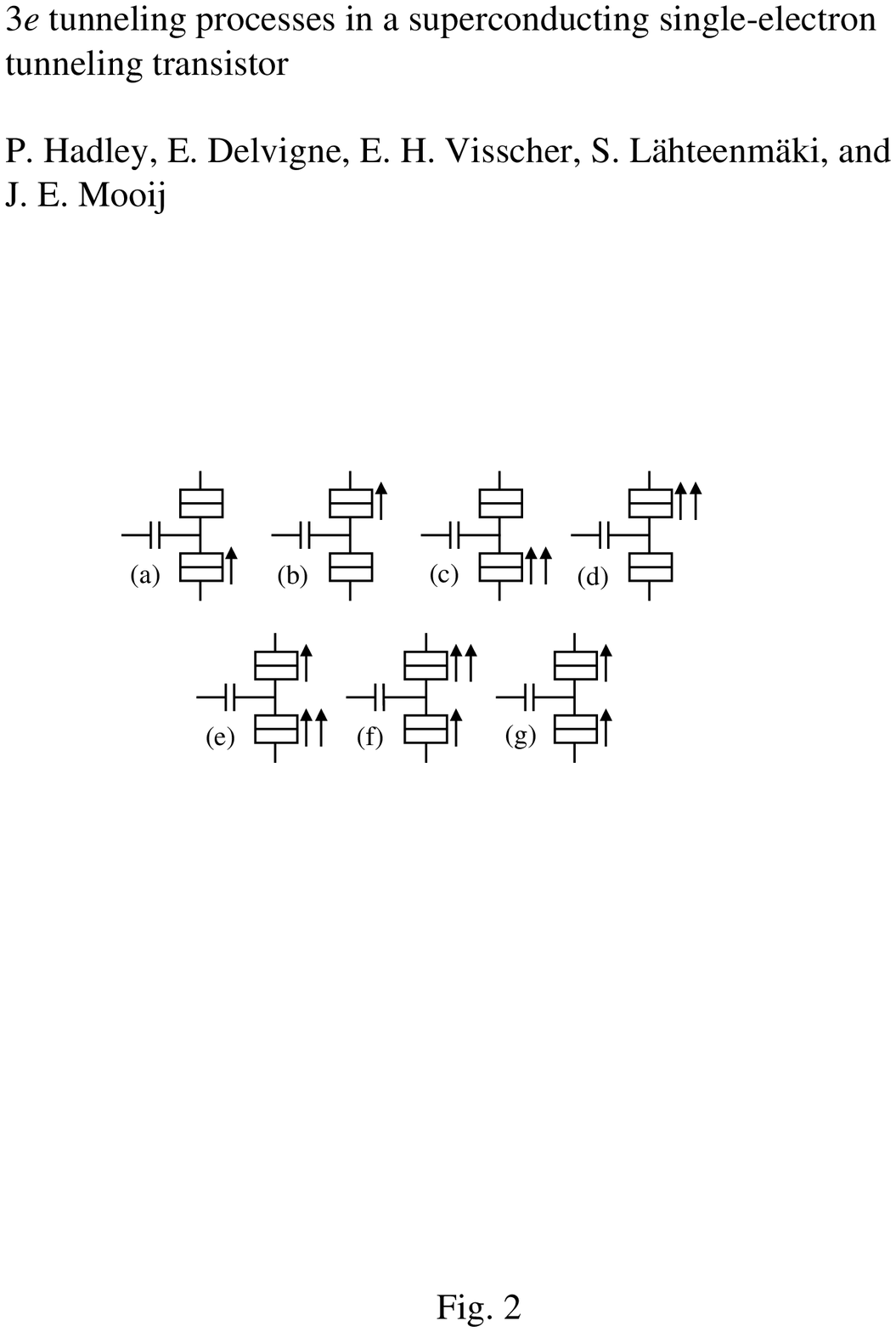, width=20.5pc, clip=true}
\caption{Nine tunnel processes were observed in the experiment. 
Each arrow indicates that a charge of $e$ has passed through 
that junction.}
\label{tunnelevents}
\end{figure}

Taking the derivatives of the electrostatic energy with 
respect to the three voltages $(V,V_{b},V_{g})$ yields a 
set of three coupled equations which can be written in the 
form
$q_i=\frac{\partial E}{\partial V_{i}}=\sum_{j}C_{ij}V_{j}$. 
Here $q_{i}$ are the charges on the islands and $C_{ij}$ is 
the capacitance matrix. The electrostatic energy of the circuit 
can then be rewritten as 
$E=\frac{1}{2}\sum_{i,j}C_{ij}^{-1}q_{i}q_{j}$. 
\cite{jackson75} This form was used to calculate the 
change in electrostatic energy when charge tunneled. 
Figure\ \ref{tunnelevents} illustrates the tunnel events 
that were considered. Each arrow indicates that a charge of 
$e$ has passed through that tunnel junction. In the limit 
$C_{b}>>C_{1},C_{2},C_{g}$, the changes in the electrostatic 
energies are,  

\begin{mathletters}
\label{electrostaticE}
\begin{equation}
\delta E=\frac{e}{C_{\Sigma }}\left[ \frac{e}{2}
-ne-q_{0}-C_{2}V_{b}-C_{g}V_{g}\right], 
\label{e1}
\end{equation}
\begin{equation}
\delta E=\frac{e}{C_{\Sigma }}\left[ \frac{e}{2}
+ne+q_{0}-\left( C_{1}+C_g \right)V_{b}+C_{g}V_{g}\right], 
\label{e2}
\end{equation}
\begin{equation}
\delta E=\frac{2e}{C_{\Sigma }}\left[ e-ne-q_{0}
-C_{2}V_{b}-C_{g}V_{g}\right], 
\label{2e1}
\end{equation}
\begin{equation}
\delta E=\frac{2e}{C_{\Sigma }}\left[ e+ne+q_{0}
-\left( C_{1}+C_g \right) V_{b}+C_{g}V_{g}\right], 
\label{2e2}
\end{equation}
\begin{equation}
\delta E=\frac{e}{C_{\Sigma }}\left[ \frac{e}{2}
-ne-q_{0}-\left( C_{1}+2C_2+C_g \right) V_{b}
-C_{g}V_{g}\right], \label{2e1e2}
\end{equation}
\begin{equation}
\delta E=\frac{e}{C_{\Sigma }}\left[ \frac{e}{2}
+ne+q_{0}-\left( 2C_{1}+C_2+2C_g \right) V_{b}+C_{g}V_{g}\right]. 
\label{e12e2}
\end{equation}
\end{mathletters}

\begin{figure} [htb]
\epsfig{figure=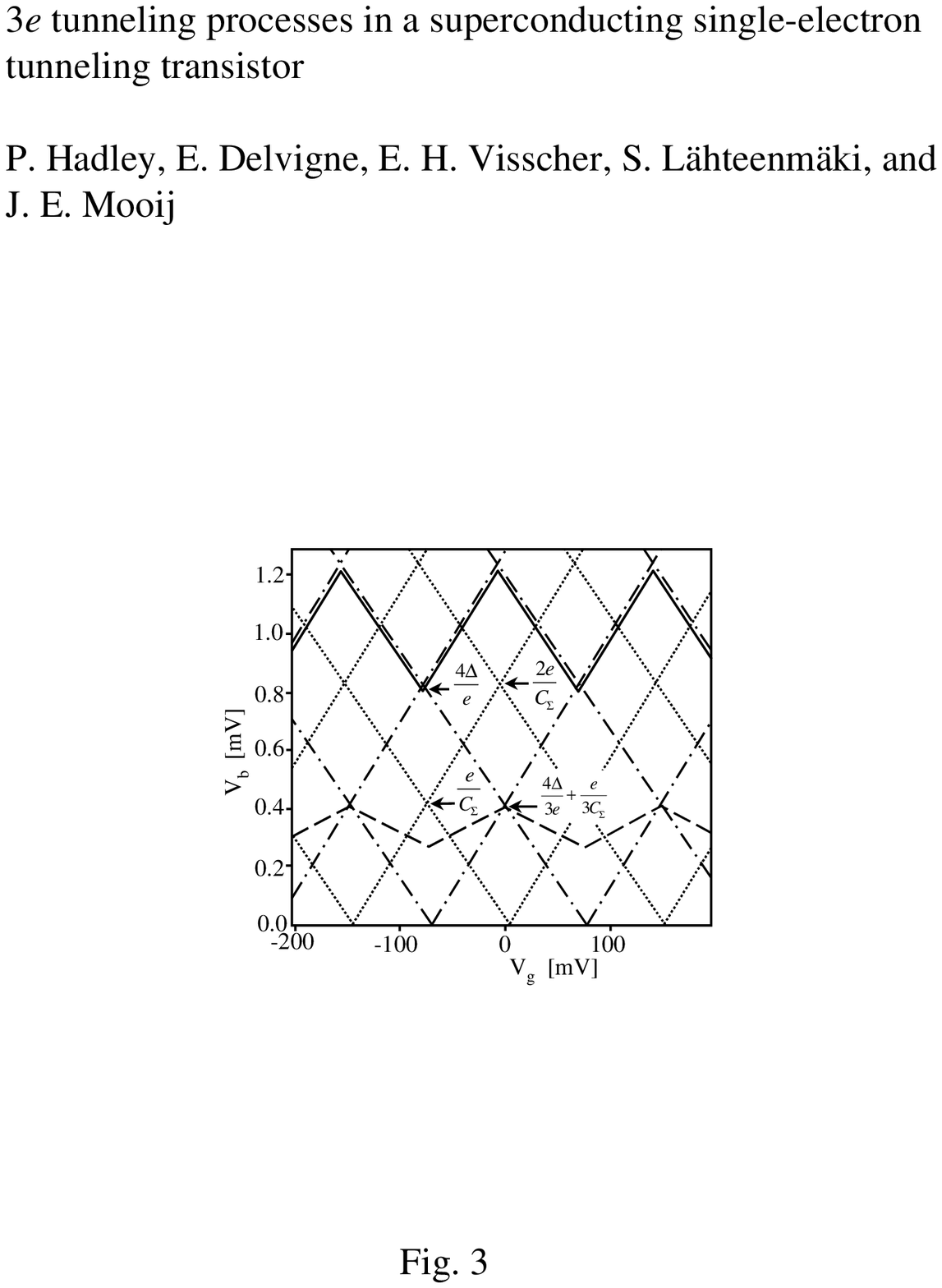, width=20.5pc, clip=true}
\caption{The thresholds for various tunnel processes in a 
superconducting SET transistor as a function of gate voltage 
and bias voltage. The solid line is the threshold for sequential 
quasiparticle tunneling (Fig. 2a and Fig.2b, $\delta E = -2 \Delta$), 
the dot - dash lines are the thresholds for singularity matching 
(Fig. 2a and Fig. 2b, $\delta E = 0$) and coincide with the 
threshold for the Coulomb blockade in the normal state. 
The dotted lines are the resonant conditions for tunnel events 
involving Cooper pair tunneling 
(Fig. 2c and Fig. 2d, $\delta E = 0$). The dashed lines are the 
thresholds for the tunneling of $3e$ of charge (Fig. 2e and Fig. 
2f, $\delta E = -2 \Delta $). The experimental values were used 
to generate this figure. The program that was used to generate 
the figure is available at 
http://vortex.tn.tudelft.nl/research/set/stability/stability.html}
\label{thresholds}
\end{figure}

Equation 2x corresponds to the tunnel event illustrated in 
Fig. 2x. The changes in electrostatic energy can be used 
to construct 
a stability diagram for the superconducting SET transistor 
as shown in Fig.\ \ref{thresholds}. Each line in 
Fig.\ \ref{thresholds} represents the threshold for a 
certain tunnel process. The position of the threshold is 
dependent on the number of electrons on the island, $n$. 
This results in a periodic stability diagram with a 
periodicity $e$. The lines which are determined by the 
tunneling of charge only through junction 1 (Fig. 2a and 
Fig. 2c) have a slope of $-C_{g}/C_{2}$. The lines which 
are determined by the tunneling of charge only through 
junction 2 (Fig. 2b and Fig. 2d) have a slope of 
$C_{g}/(C_{1}+C_{g})$. The threshold

determined by the 
tunneling of 3$e$ of charge as shown in Fig. 2e has a slope 
of $-C_{g}/(C_{1}+2C_{2}+C_{g})$ and the slope of the threshold 
determined by the tunnel process shown in Fig. 2f is 
$C_{g}/(2C_{1}+C_{2}+2C_{g})$.
	
Figure 4a shows the measured current through 
the superconducting SET transistor as a function of the bias 
voltage and the gate voltage. The logarithm of the current 
was taken so that the high bias data and low bias
\begin{figure} [htb]
\epsfig{figure=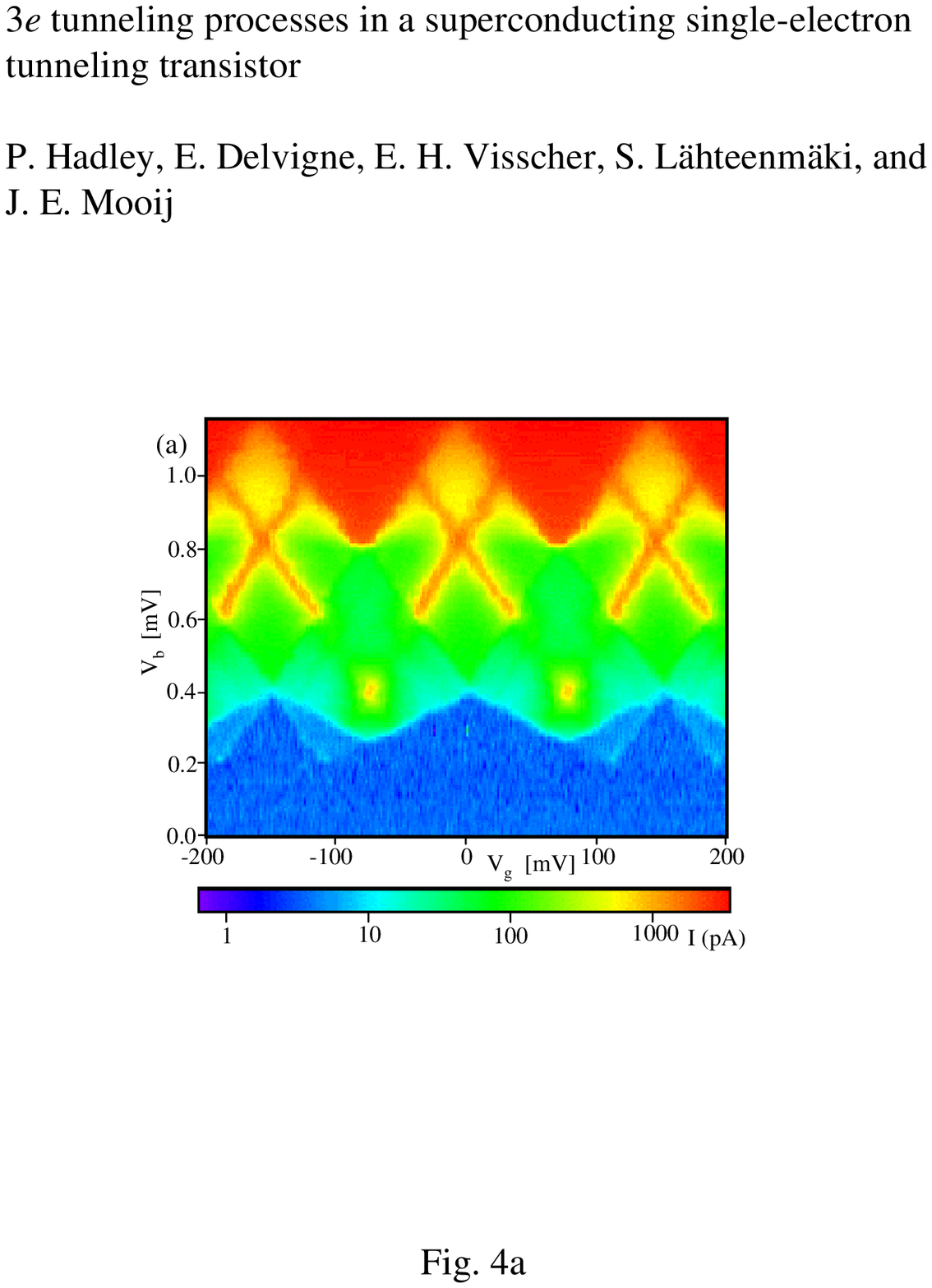, width=20.5pc, clip=true}
\epsfig{figure=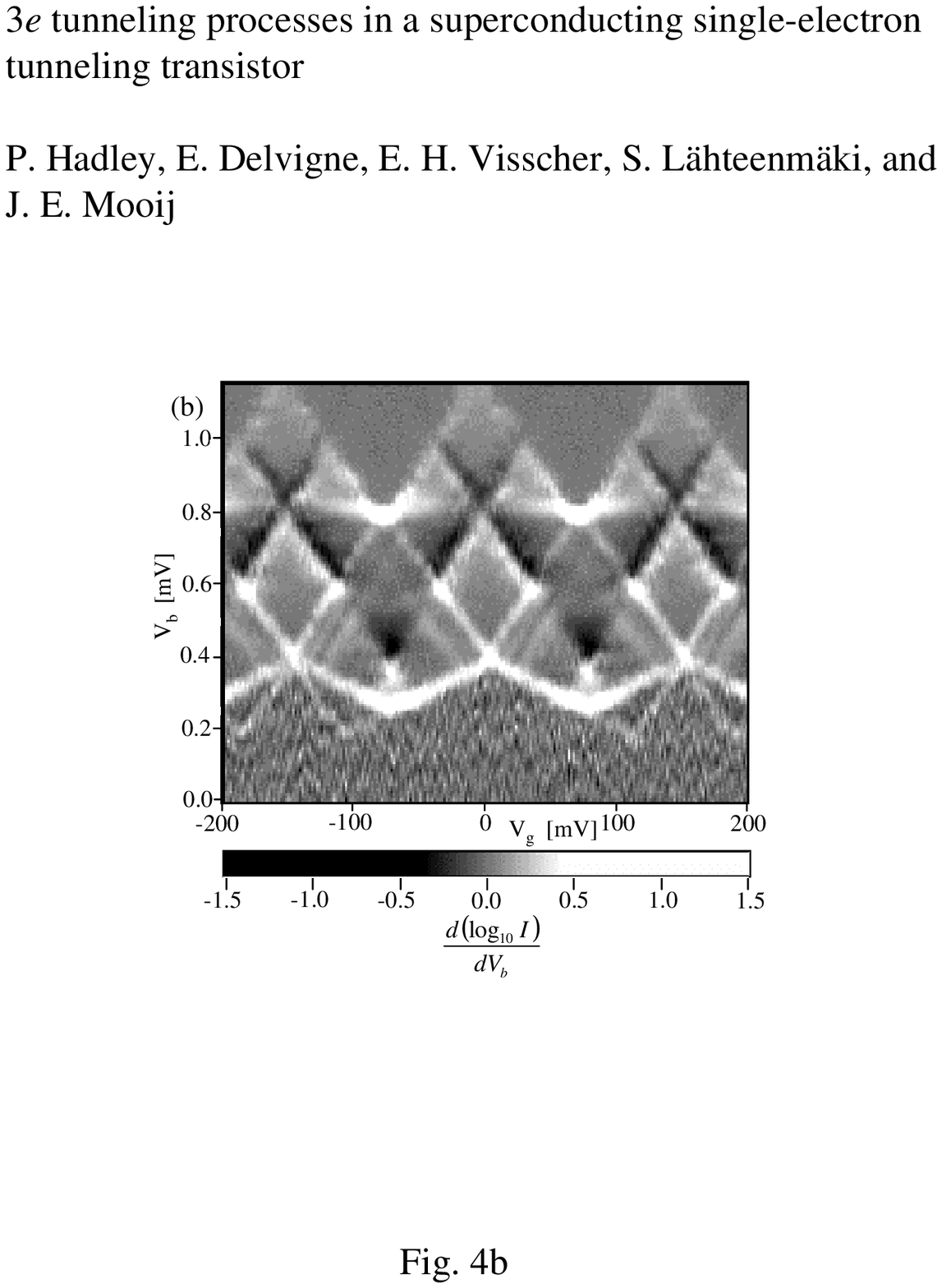, width=20.5pc, clip=true}
\caption{(a) The logarithm of the current through a 
superconducting SET transistor is plotted as a 
function of the bias voltage and the gate voltage.
(b) The same derivative of the same data shown in (a).}
\label{experiment}
\end{figure} 
data could 
be presented in the same figure. Figure 4b shows the derivative 
of the same data. The current is periodic in 
gate voltage with a periodicity of $e/C_{g}$. The inverted 
white triangles at the top of Fig. 4a form the 
threshold for sequential quasiparticle tunneling through the 
SET transistor. In this process, a single quasiparticle tunnels 
onto the island through one junction and then another 
quasiparticle tunnels out through the other junction. These 
tunnel processes are shown in Fig. 2a ($\delta E=-2\Delta $) 
and Fig. 2b 
($\delta E=-2\Delta $). The minimum bias voltage for the 
threshold for sequential quasiparticle tunneling is 
$4\Delta /e$ and the maximum is $4\Delta /e+e/C_{\Sigma }$. 
Here $C_{\Sigma }$ is the total capacitance, 
$C_{\Sigma }=C_{1}+C_{2}+C_{g}$. The change in electrostatic 
energy when a quasiparticle tunnels must be 
$\delta E=-2\Delta $ because there are no quasiparticle states 
within the superconducting gap. 

Also clearly visible in Fig. 4a are intersecting 
ridges of current that are due to the Josephson - quasiparticle 
(JQP) cycle. These are the white Xs centered at about 0.8 mV. 
This transport mechanism can occur when the bias voltages are such
that a Cooper pair can be transported through one of the junctions
without changing the total energy of the system. There 
are then two degenerate charge states which are coupled by the 
Josephson energy $E_J$. This results in a mixing of the charge states 
and the probability of the Cooper pair being on either side of the 
junction
oscillates with a frequency $E_J/\hbar=I_c/(2e)$. These oscillations
produce no net current, however the oscillations can be interrupted 
by the tunneling of a quasiparticle through the other junction. The 
result of this interruption is that a Cooper pair is transported
through one of the junctions while a quasiparticle is
transported through the other junction. The charge of 
the island changes by $e$, and the mixing of the charge states ceases. 
If the bias voltage is greater than $2\Delta /e + e/C_{\Sigma}$, then a 
second quasiparticle can tunnel returning the system to its original 
charge state and the process
can start over again. The JQP current ridges intersect at a bias 
voltage of $2e/C_{\Sigma}$.  

There are also isolated current peaks located at a bias of 0.4 
meV in Fig. 4. These peaks lie on the 
extensions of the JQP current ridges at a bias voltage of $e/ C_{\Sigma}$. 
\cite{brink91,siewert96} Two sequential tunneling events are 
responsible for these current peaks that are similar to the first tunnel 
process in the JQP cycle described above.  First Cooper pair tunneling is
resonant across junction 1. When the tunneling of a quasiparticle
through junction 2 interrupts the mixing of the charge states, a charge 
of $-2e$ is transported through junction 1 and a charge of $-e$ is 
transported through junction 2. This adjusts the potential of the island
so that Cooper pair tunneling is resonant across junction 2. Then a
quasiparticle can tunnel onto the island through junction 1 while a 
Cooper pair is transported off the island through junction 2. This 
returns the system to its original charge state and the process 
repeats.   

The horizontal line at $4\Delta$ in Fig. 4
is due to the rather abrupt onset of cotunneling of 
quasiparticles at a bias voltage of $4\Delta$. This cotunneling 
is illustrated in Fig. 2g. Cotunneling of quasiparticles for 
bias voltages less than $4\Delta$ is suppressed by the lack 
of quasiparticle states in the superconducting gap.\cite{averin97}  

Now we focus on the sawtooth threshold for current that 
lies just below 0.4 mV in Fig. 4. This 
threshold is $e$ periodic and the lines that form the 
threshold have a slope that is one third of the slope of 
the threshold for sequential quasiparticle tunneling or the 
JQP cycle. The tunnel process responsible for this threshold 
is one where a Cooper pair and a quasiparticle tunnel 
simultaneously. This sort of cotunneling event involving a 
Cooper pair and a quasiparticle was first described by 
Maassen van den Brink et al. \cite{brink291} First the 
charge on the island decreases by $-e$ via the tunnel event 
shown in Fig. 2e with $\delta E=-2\Delta $. Then the island 
returns to its initial charge state via the tunnel event in 
Fig. 2f with $\delta E=-2\Delta $. The minimum bias voltage 
for this threshold is $4 \Delta /(3e)$ and the maximum bias 
voltage for this threshold is for this process is 
$4 \Delta /(3e) + e/(3C_{\Sigma})$. A similar simultaneous 
$3e$ tunneling threshold should also occur for SET transistors 
in the normal state (Fig. 2e and Fig. 2f, $\delta E=0$). 
However in that case three particles would have to tunnel 
simultaneously so the rate would be much lower. 

The tunneling of $3e$ of charge also forms part of a sequence 
of tunnel events that is responsible for the current observed 
in the diamond shaped regions that extend from a bias voltage 
of about 0.4 mV to 1.2 mV. In this region, first $3e$ of 
charge tunnels as in Fig. 2e (or Fig. 2f) with 
$\delta E = -2\Delta $. Then the charge of the island returns 
to its initial state by the tunneling of a quasiparticle as 
in Fig. 2b (or Fig. 2a) with $\delta E = -2\Delta $.

At bias voltages between 0.2 mV and 0.4 mV a small current that 
is $2e$ periodic is observed. This current arises from the 
sequential tunneling of a quasiparticle and the tunneling of 
$3e$ of charge as described above. If the initial state of 
the island is odd, then a quasiparticle can tunnel on or off 
the island in the tunnel processes illustrated in Fig. 2a or 
Fig. 2b with $\delta E=0$.\cite{schoen94} In this tunnel process, 
the quasiparticle that tunnels pairs with the odd quasiparticle on 
the island. The island then returns to its initial charge state 
via a Cooper pair-quasiparticle cotunneling event (Fig. 2e or 
Fig. 2f, $\delta E=-2\Delta $). A similar process cannot occur 
if the initial state of the island is even since the 
quasiparticle that tunnels from the lead has no partner to 
condense with to form a Cooper pair. Consequently, this current 
is $2e$ periodic. 

In summary, the thresholds for a number of distinct charge 
transport mechanisms were observed in the current-voltage 
characteristics of a superconducting SET transistor. These 
cycles involve the sequential tunneling of quasiparticles, 
the sequential tunneling of Cooper pairs and quasiparticles 
(JQP Cycles), cotunneling of quasiparticles, and the sequential 
cotunneling of Cooper pairs and quasiparticles with the 
tunneling of quasiparticles. Of particular interest are 
the currents that arise from cycles which include cotunneling 
of a Cooper pair and a quasiparticle. In this tunnel process, 
a charge of $3e$ tunnels and the Cooper pair and quasiparticle 
are transported simultaneously through two different tunnel 
barriers. Cotunneling of a Cooper pair and a quasiparticle 
also plays a role in a sequence of tunnel events that leads 
to a $2e$ periodic current at low bias voltages.

\acknowledgments

We are indebted to Sarah Pohlen, Leonid Glazman, Yuli Nazarov, 
Gerd Sch\"on, and Arkadi Odintsov their enlightening comments 
on this work. We also thank Caspar van der Wal for assistance
with the measurements.
Support from Esprit project 22953, CHARGE, is 
gratefully acknowledged.

\widetext %need to end this multicol stuff before the end of the document
\end{document}